\documentclass[11pt,a4paper,cite]{article}
\usepackage[english]{babel}
\usepackage[T1]{fontenc}
\usepackage[latin1]{inputenc}
\usepackage{epsfig}
\usepackage{graphics}
\usepackage{cite}
\usepackage{amsmath}
\usepackage{amsfonts}
\usepackage{amssymb}
\usepackage{amsthm}
\usepackage[makeroom]{cancel}
\usepackage{color}
\usepackage{bm}
\usepackage{datetime}
\usepackage{xcolor}

\newcommand{\beq}{\begin{equation}}
\newcommand{\eeq}{\end{equation}}

\newcommand\blfootnote[1]{%
	\begingroup
	\renewcommand\thefootnote{}\footnote{#1}%
	\addtocounter{footnote}{-1}%
	\endgroup
}

\newdateformat{monthyeardate}{%
	\monthname[\THEMONTH], \THEYEAR}

\title{\bf On the inexistence of  self-gravitating solitons\\  in generalised axion electrodynamics}
\author{Carlos A. R. Herdeiro$^{1,2\ddagger}$, Jo\~{a}o M. S. Oliveira$^{2\dagger}$,}
\date{%
	$^1${\small Centro de Astrof\'\i sica e Gravita\c c\~ao - CENTRA,} \\ {\small Departamento de F\'\i sica,
Instituto Superior T\'ecnico - IST, Universidade de Lisboa,} \\  {\small Avenida
Rovisco Pais 1, 1049-001, Portugal}\\ \vspace{0.3cm}
$^2${\small Departamento de Matem\'atica da Universidade de Aveiro and CIDMA,} \\ {\small Campus de Santiago, 3810-183 Aveiro, Portugal}\\
 \vspace{0.3cm}
	\monthyeardate\today
}

\begin{document}
\maketitle
\blfootnote{$\ddagger$ herdeiro@ua.pt}
\blfootnote{$\dagger$ jmiguel.oliveira@ua.pt}

\begin{abstract}
Building upon the methods used recently in \cite{HerdeiroOliveira}, we establish the inexistence of self-gravitating  solitonic solutions for both static and strictly stationary asymptotically flat spacetimes in generalised axion electrodynamics. This is an Einstein-Maxwell-axion model, where the axion field $\theta$ is non-minimally coupled to the electromagnetic field. Considering the standard QCD axion coupling, we first present an argument for the absence of static axionic solitons, $i.e.$ localised energy axionic-electromagnetic configurations, yielding an everywhere regular, horizonless, asymptotically flat, static spacetime. Then, for generic couplings $f(\theta)$ and $g(\theta)$ (subject to mild assumptions) between the axion field and the electromagnetic field invariants, we show there are still no solitonic solutions, even when dropping the staticity assumption and merely requiring a strictly stationary spacetime, regardless of the spatial isometries.
\end{abstract}

\newpage

\section{Introduction}
Quantum Chromodynamics (QCD) admits a term that violates the combined CP (Charge conjugation and Parity) discrete symmetries. Yet, such violation is not observed in any experimental process controlled by the strong interaction only, which suggests that if it exists it must be very small. Consequently, the CP violating term must have an unnaturally small coefficient, yielding a fine-tuning problem. 

An ingenious solution to this \textit{strong CP problem} was proposed by Peccei and Quinn \cite{PecceiQuinn1, PecceiQuinn2}. Their idea was to promote the unnaturally small coefficient into a dynamical field which could be dynamically relaxed to zero. In the original guise, the mechanism extended the Standard Model with a complex scalar field possessing a global $U(1)$ symmetry and a Mexican hat type potential. The symmetry is spontaneously broken below some high energy scale, wherein the complex scalar acquires a vacuum expectation value (vev), yielding a  Goldstone pseudo-scalar -- the \textit{axion}, $\theta(x)$ -- parameterising the degeneracy of the  potential vacuum manifold. If, moreover, at least one of the fermions in the model acquires its mass via a Yukawa coupling to the complex scalar, the axion acquires a potential under a chiral anomaly, driving it to a vev that precisely cancels the CP violating term and, moreover, endows the axion with a small mass \cite{AxionWeinberg, AxionWilczek}. When later studied in a cosmological context, it was suggested that axions are interesting dark matter candidates \cite{AxDM1, AxDM2, AxDM3}, see also \cite{AxDMR}. Since then, the study of gravitational effects of axion-like particles have received considerable attention.

In this work we discuss whether \emph{solitonic} self-gravitating solutions,  i.e., everywhere non-singular,  asymptotically flat spacetimes without a horizon, are possible in a generalised axion electrodynamics minimally coupled to Einstein's gravity -  Einstein-Maxwell-axion models. Such solitons would describe localised lumps of energy, particle-like solutions, in a field theory-gravity model. To contextualise this question, it  is known that  the Einstein-Maxwell system admits no static soliton solutions -  see $e.g.$~ \cite{Heusler96}. The same holds for strictly stationary,  but not necessarily static, spacetimes \cite{Shiromizu:2012hb}. Similar conclusions still hold if one considers Einstein-Maxwell-scalar models~\cite{HerdeiroOliveira} without axion-like couplings, but allowing generic couplings between the scalar field and the Maxwell invariant. Thus, one may ask whether the particle physics motivated axion coupling to the electromagnetic field \cite{SikivieModel} could change this state of affairs, possibly unveiling another guise for axionic manifestation in Nature. 

The possible existence of self-gravitating axionic solitons  is discussed here for both static and strictly stationary configurations. Our analysis starts with the simplest Einstein-Maxwell-axion model, but a generalisation, considering a non-minimal coupling $f(\theta)$ between the axion field and the Maxwell term, is also discussed, following~\cite{HerdeiroOliveira}. 

This paper is organised as follows: in section~\ref{section2} we discuss the absence of self-gravitating, static axionic solitons,  with the usual axion electrodynamics description. In section~\ref{section3}, we include a non-minimal coupling $f(\theta)$ between the axion field and the electromagnetic field and change the usual linear axion coupling $\kappa\theta$ to an arbitrary pseudoscalar function $g(\theta)$, finding that a no-go theorem for solitons still holds. In section~\ref{section4} we generalise the latter result for strictly stationary, but not necessarily static, configurations. Finally, in section~\eqref{section5}, we present a discussion of the results and possible future work.

\section{Absence of static axionic solitons}
\label{section2}

Axion electrodynamics minimally coupled to Einstein's gravity consists, apart from the gravitational action, of the usual Maxwell and Klein-Gordon terms along with an additional term which couples the electromagnetic field to the axion field. It is represented by the following action\cite{SikivieModel,AxModel1,AxModel2}
\beq\label{ActionAx}
\mathcal{S}_{Ax} = \mathcal{S}_{EH} + \int d^4x\sqrt{-g}\bigg[- \frac{1}{4}F_{\mu\nu} F^{\mu\nu} + \frac{\kappa\theta}{4}F_{\mu\nu}\tilde{F}^{\mu\nu} + \frac{1}{2}\nabla_\mu\theta\nabla^\mu\theta - U(\theta)\bigg] \ ,
\eeq
where $\mathcal{S}_{EH} $ is the Einstein-Hilbert action, $\kappa$ is a constant, $\theta$ is the pseudo-scalar axion field and $\tilde{F}^{\mu\nu}$ is the Hodge dual of the Maxwell tensor $F_{\mu\nu}=\partial_\mu A_\nu-\partial_\nu A_\mu$, $\tilde{F}^{\mu\nu}=\frac{1}{2}\epsilon^{\mu\nu\alpha\beta}F_{\alpha\beta}$, where $\epsilon^{\mu\nu\alpha\beta}$ is the contravariant Levi-Civita tensor. We also allow for the existence of a general axion potential $U(\theta)$. In this section we consider an asymptotically flat, static spacetime with no restrictions on the spatial symmetries. The gravitational part will play no role in the subsequent argument.

The equations of motion for this model are
\begin{align}
\nabla_\mu(F^{\mu\nu}& - \kappa\theta \tilde{F}^{\mu\nu}) = 0 \ ,\label{EqF}\\
\nabla_\mu\tilde{F}^{\mu\nu}& = 0 \ , \\
\Box \theta = \frac{\kappa}{4}&F_{\mu\nu}\tilde{F}^{\mu\nu}  -\frac{d U(\theta)}{d\theta} \ , \label{EqHF}
\end{align}
where $\Box$ is the covariant d'Alembertian. Since the spacetime is static and without horizons it admits an everywhere timelike Killing vector field $k$. This vector field can be used to define the electric and magnetic fields (in fact 4-(co)vectors) as:
\begin{align}
E_\mu &\equiv  -F_{\mu\nu}k^{\nu} \label{efield}\ , \\
B_\mu &\equiv -\frac{1}{2}\varepsilon_{\mu\alpha\beta\nu} F^{\alpha\beta} k^\nu = -\tilde{F}_{\mu\nu}k^{\nu} \label{bfield} \ .
\end{align}

In Maxwell's theory, one can rewrite the covariant Maxwell equations in terms of $E,B$ in a certain canonical form - see $e.g.$ eqs. (38)-(41) in~\cite{HerdeiroOliveira}. In axion electrodynamics, a similar canonical form is obtained if we define two new fields $E'$ and $B'$ which are related to the original fields as
\begin{align}
E'_\mu &\equiv  E_\mu - \kappa\theta B_\mu \label{Eredef} \ ,\\
B'_\mu & \equiv  B_\mu + \kappa\theta E_\mu \label{Bredef} \ ;
\end{align}
now, the axion Maxwell equations~\eqref{EqF}-\eqref{EqHF} are written as 
\begin{align}
&\nabla_{[\mu}E_{\nu]}= 0\ , \label{MaxA1} \\
&\nabla_{[\mu}B'_{\nu]} = 0 \ , \label{MaxA2} \\
&\nabla_\mu\left(\frac{E'^\mu}{V} \right) = 0 \ , \label{MaxA3}\\
&\nabla_\mu\left(\frac{B^\mu}{V} \right) = 0 \ , \label{MaxA4}
\end{align}
where $V\equiv -k^\mu k_\mu>0$. Due to the absence of currents, the first two equations imply that an electric $\varphi$ and a magnetic $\psi$ \textit{scalar} potentials can be introduced, as 
\begin{equation}
E_\mu=\partial_\mu\varphi \ , \qquad   B'_\mu=\partial_\mu\psi \ .
\label{potentials}
\end{equation}

The remainder of the argument uses the method in \cite{HerdeiroOliveira} which was inspired by Heusler's argument described in \cite{Heusler96}. We make use of the following identity: for an arbitrary vector $\alpha$ obeying $\pounds_k\alpha=[k,\alpha]=0$, it holds that~\cite{HeuslerBook}:
\beq
\int_{\partial\Sigma}\alpha^{\mu}k^{\nu}dS_{\mu\nu} = \frac{1}{2}\int_{\Sigma} \nabla_\mu\alpha^\mu k^\nu d\Sigma_\nu\label{Stokes} \ ,
\eeq
where $\Sigma$ is an arbitrary Cauchy hypersurface with volume element $d\Sigma_\nu$ and boundary $\partial\Sigma$, the latter with antisymmetric area element $dS_{\mu\nu}$. Specifying this identity for $\alpha^\mu = E'^\mu/V$ and using the axionic Maxwell equations yields
\beq
\int_{\partial\Sigma}\frac{E'^{\mu}k^{\nu}}{V}dS_{\mu\nu} = 0\ ,
\label{int0}
\eeq
where we took $\partial\Sigma$ to be the surface at spatial infinity (an $r=\infty$ 2-surface, where $r$ is the standard Minkowski radial coordinate, which can be used near infinity due to asymptotic flatness). 

Making a second use of the identity~\eqref{Stokes} but now with $\alpha^\mu=\varphi E'^\mu/V$ and once again using the axionic equations, we obtain
\beq
\frac{1}{2}\int_{\Sigma}\frac{E^\mu E'_\mu}{V}k^\nu d\Sigma_\nu = \int_{\partial\Sigma}\varphi \frac{E'^{\mu}k^{\nu}}{V} dS_{\mu\nu} = \varphi_\infty\int_{\partial\Sigma} \frac{E'^{\mu}k^{\nu}}{V} dS_{\mu\nu} = 0 \ ,
\eeq
where $\varphi_\infty$ is the value of the electric potential at $r=\infty$ which is constant, and the last equality used~\eqref{int0}.

The same argument can be used for $B$ and $B'$ by replacing $\varphi$ by $\psi$, obtaining
\beq
\int_{\Sigma}\frac{B^\mu B'_\mu}{V}k^\nu d\Sigma_\nu = 0 \ .
\eeq
We can now expand $(E',B')$ in terms of $(E,B)$, via~\eqref{Eredef}-\eqref{Bredef} to obtain the identities:
\begin{align}
\int_{\Sigma}\frac{E^\mu E_\mu}{V}k^\nu d\Sigma_\nu -\int_{\Sigma}\kappa\theta\frac{E^\mu B_\mu}{V}k^\nu d\Sigma_\nu &= 0 \ ,
\label{identities1} \\
\int_{\Sigma}\frac{B^\mu B_\mu}{V}k^\nu d\Sigma_\nu +\int_{\Sigma}\kappa\theta\frac{E^\mu B_\mu}{V}k^\nu d\Sigma_\nu &= 0 \ .
\label{identities2}
\end{align}
Adding up the last two equations yields
\beq
\int_{\Sigma}\frac{E^\mu E_\mu + B^\mu B_\mu}{V}k^\nu d\Sigma_\nu = 0 \ .
\label{cons}
\eeq

From their definitions~\eqref{efield}-\eqref{bfield}, $k^\mu E_\mu = 0 = k^\mu B_\mu$. Thus, these fields are never timelike. It follows that both $E^\mu E_\mu$ and $B^\mu B_\mu$ are always non-negative. Consequently, the only way for eq.~\eqref{cons} to be verified is if both fields vanish for every Cauchy surface $\Sigma$ and, consequently, for the whole spacetime. This result is independent of the potential $U(\theta)$. With vanishing electromagnetic fields, all we have left is the possibility of self-gravitating axion (scalar) solitons. However it has been shown that there are no scalar field solitons as long as the dominant energy condition is obeyed and the strong energy condition is violated, which is the case for scalar fields with a positive potential (see \cite{HerdeiroOliveira, HeuslerBook, Heusler95}). Therefore, the only possible solution for such potentials is Minkowski spacetime.

As a final remark in this section, the main difference between the result herein and the one for Einstein-Maxwell theory is that instead of establishing that the norms of both $E$ and $B$ vanish, we can only establish that the sum of these norms must vanish. Since both these norms are positive definite, however, the final conclusion is that each must vanish, recovering the result of Einstein-Maxwell theory.

\section{Generalised axion electrodynamics}
\label{section3}
The result of section~\eqref{section2} can be straightforwardly extended to a model of generalised axion electrodynamics minimally coupled to Einstein's gravity
\beq\label{ActionAxA}
\mathcal{S}_{A} = \mathcal{S}_{EH} + \int d^4x\sqrt{-g}\bigg[- \frac{f(\theta)}{4}F_{\mu\nu} F^{\mu\nu} + \frac{g(\theta)}{4}F_{\mu\nu}\tilde{F}^{\mu\nu} + \frac{1}{2}\nabla_\mu\theta\nabla^\mu\theta - U(\theta)\bigg] \ ,
\eeq
which introduces the arbitrary functions $f(\theta)$ and $g(\theta)$ of the axion field. The function $g(\theta)$ is a pseudoscalar function and $f(\theta)$ is a non-minimal coupling between the axion and the standard Maxwell  term, as discussed in \cite{HerdeiroOliveira} motivated by the recent results of scalarisation in Einstein-Maxwell-scalar models~\cite{Herdeiro:2018wub}. 

In order to recover Einstein-Maxwell when there is no axion, we assume that $f(0)=1$. It is also assumed that both functions do not diverge in our spacetime\footnote{This assumption is considered as the application of the Stokes theorem would include constant contributions due to divergences in the spacetime. Therefore, our approach is not valid for diverging coefficient functions.}. The equations of motion are a simple generalisation of the previous ones \eqref{EqF}-\eqref{EqHF} and read
\begin{align}
\nabla_\mu(fF^{\mu\nu}& - g \tilde{F}^{\mu\nu}) = 0 \ ,\\
\nabla_\mu\tilde{F}^{\mu\nu}&=0 \ ,\\
\Box^2\theta = \frac{1}{4}&\frac{d g}{d\theta}F_{\mu\nu}\tilde{F}^{\mu\nu}  - \frac{1}{4}\frac{d f}{d\theta}F_{\mu\nu}F^{\mu\nu} -\frac{d U(\theta)}{d\theta} \ .
\end{align}
Although the $\theta$ equation can be considerably more difficult due to the arbitrary couplings, defining now the fields $E'$ and $B'$ as
\begin{align}
E' &= fE - gB \ ,\label{newE}\\
B' &= fB + gE \ , \label{newB}
\end{align}
it follows that these new fields respect the exact same equations as \eqref{MaxA1}-\eqref{MaxA4}. Consequently,  we follow the exact same procedure as in section~\ref{section2} to obtain the corresponding relations to~\eqref{identities1}-\eqref{identities2}, which now read
\begin{align}
\int_{\Sigma}f\frac{E^\mu E_\mu}{V}k^\nu d\Sigma_\nu -\int_{\Sigma}g\frac{E^\mu B_\mu}{V}k^\nu d\Sigma_\nu &= 0 \ ,\label{g1} \\
\int_{\Sigma}f\frac{B^\mu B_\mu}{V}k^\nu d\Sigma_\nu +\int_{\Sigma}g\frac{E^\mu B_\mu}{V}k^\nu d\Sigma_\nu &= 0 \ .\label{g2}
\end{align}
Adding these equations now yields
\beq
\int_{\Sigma}f\frac{E^\mu E_\mu + B^\mu B_\mu}{V}k^\nu d\Sigma_\nu =0 \ .
\eeq
As both $E^\mu E_\mu$ and $B^\mu B_\mu$ are non-negative,  this identity implies a similar result to the one obtained in \cite{HerdeiroOliveira} for the theory with no axions ($g=0$): the fields must vanish and there are no solitonic solutions as long as the coupling $f(\theta)$ does not change sign. We can see that the main reason for this result to be similar to the one with $g=0$ is because $g$, as complicated a function as it might be, does not contribute to the argument due to its contribution disappearing when we add equations \eqref{g1} and \eqref{g2}.

\section{Absence of strictly stationary axionic solitons}
\label{section4}

So far we have considered static spacetimes. The method used above allowed us to rule out static solitons without requiring any spatial isometry (see~\cite{Gibbons:1990um} for other approaches to establish the absence of static solitons).
Now we wish to consider strictly stationary, but not necessarily static, axionic solitons with the more general model \eqref{ActionAxA}. This accounts now for possibly rotating solitons, as long as rotation does not create ergo-regions, since strict stationarity means that there exists an \textit{everywhere} timelike Killing vector field.  Following a procedure similar to \cite{HerdeiroOliveira} where we use a Lichnerowicz type argument, see $e.g.$\cite{Shiromizu:2012hb}, we shall also establish a no-go theorem for solitons. In this case the Einstein equations play an important role in the argument.

The Einstein equations for this model are
\beq\label{EinsteinEq}
R_{\mu\nu} = f(\Phi)\left(F_\mu^{\,\,\alpha}F_{\nu\alpha} - \frac{1}{4}g_{\mu\nu}F^2\right) + \partial_\mu\Phi\partial_\nu\Phi + g_{\mu\nu}U(\Phi) \ .
\eeq
The axionic term is purely topological so it does not contribute to the Einstein equations.
Using the timelike Killing vector field, we define the \textit{twist vector} $\omega^\mu$ as
\beq
\omega^\mu = \frac{1}{2}\varepsilon^{\mu\nu\alpha\beta}k_\nu\nabla_\alpha k_\beta \ ;
\label{twistvector}
\eeq
this vector obeys
\beq\label{TwistId}
\nabla_\mu\bigg(\frac{\omega^\mu}{V^2}\bigg)=0 \ .
\eeq
The Maxwell equations ~\eqref{MaxA1}-\eqref{MaxA4}, with the primed fields defined by~\eqref{newE}-\eqref{newB} are generalised for a strictly stationary spacetime as:
\begin{align}
&\nabla_{[\mu}E_{\nu]}= 0\ , \label{MaxA1S} \\
&\nabla_{[\mu}B'_{\nu]} = 0 \ , \label{MaxA2S} \\
&\nabla_\mu\left(\frac{E'^\mu}{V} \right) =  \frac{2}{V^2}\omega_\mu B'^\mu \ , \label{MaxA3S}\\
&\nabla_\mu\left(\frac{B^\mu}{V} \right) = -\frac{2}{V^2}\omega_\mu E^\mu \ . \label{MaxA4S}
\end{align}
It can be shown that
\beq
\nabla_{[\mu}\omega_{\nu]} = \frac{1}{2}\varepsilon_{\mu\nu}^{\;\;\;\;\alpha\beta}k_{[\alpha}R_{\beta]\gamma}k^\gamma \ ,
\eeq
so that using the Einstein equations \eqref{EinsteinEq} relates the curl of $\omega$ with the Poynting vector:
\beq\label{omegaBE}
\nabla_{[\mu}\omega_{\nu]} =  f B_{[\mu}E_{\nu]}  \ .
\eeq
One can freely add vanishing terms such as $-g B_{[\mu}B_{\nu]}$ and $g E_{[\mu}E_{\nu]}$ to rewrite the right hand side in two different ways
\beq
f B_{[\mu}E_{\nu]} =  B'_{[\mu}E_{\nu]} = B_{[\mu}E'_{\nu]}  \ .
\eeq
We choose the expression with $B'$ and $E$ as these two fields are the ones which we can rewrite as potentials $\psi$ and $\phi$ respectively, $cf.$~\eqref{potentials}. This means that equation \eqref{omegaBE} implies the following two identities
\begin{align}
\nabla_{[\mu}\left(\omega_{\nu]}-\psi E_{\nu]}\right) = 0 \ ,\\
\nabla_{[\mu}\left(\omega_{\nu]}+ \phi B'_{\nu]}\right) = 0 \ ,
\end{align}
which in turn imply the existence of two new potentials $U_{B'}$ and $U_E$
\begin{align}
\nabla_\mu U_E &= \omega_\mu-\psi E_\mu \ ,\\
\nabla_\mu U_{B'} &= \omega_\mu+ \phi B'_\mu \ .
\end{align}
Using these potentials and the identity \eqref{TwistId}, the following divergence identity is obtained
\beq\label{DivId}
\nabla_\mu W^\mu = \frac{4\omega^\mu\omega_\mu}{V^2}-\frac{E'_\mu E^\mu + B'_\mu B^\mu}{V} \ ,
\eeq
where
\beq
W^\mu = 2(U_E + U_{B'})\frac{\omega^\mu}{V^2} -\frac{\psi B^\mu+\phi E'^\mu}{V}\ .
\eeq

On the other hand, the contraction of the Einstein equations \eqref{EinsteinEq} with the Killing field yields
\beq
\label{RiccKilling}
\frac{2}{V}R_{\mu\nu}k^\mu k^\nu = f\frac{E_\mu E^\mu + B_\mu B^\mu}{V} - 2U(\theta) \ .
\eeq
The first term on the right hand side can be slightly reshaped by noting that $f(E_\mu E^\mu + B_\mu B^\mu)$ may be written as
\begin{align}
f(E_\mu E^\mu + B_\mu B^\mu) 
=(fE_\mu - g B_\mu) E^\mu + (fB_\mu + g E_\mu) B^\mu 
=E'_\mu E^\mu + B'_\mu B^\mu \ .
\end{align}
Then adding equations \eqref{DivId} and \eqref{RiccKilling} yields
\beq\label{RiccDiv}
\frac{2}{V}R_{\mu\nu}k^\mu k^\nu - \frac{4\omega^\mu\omega_\mu}{V^2} = - \nabla_\mu W^\mu - 2U(\theta) \ .
\eeq

The final step of the argument consists on taking the Komar mass integral on a Cauchy surface $\Sigma$~\cite{Heusler1995a}:
\beq
\label{KomarM}
M= -\int_\Sigma\left(\frac{2R_{\mu\nu}k^\mu k^\nu}{V} - \frac{4\omega^\mu\omega_\mu}{V^2}\right)k^\alpha d\Sigma_\alpha \ ,
\eeq
which, via \eqref{RiccDiv}, reads
\beq
M = \int_\Sigma\left( \nabla_\mu W^\mu + 2U \right)k^\alpha d\Sigma_\alpha \ .
\label{mass}
\eeq
As $\pounds_k W = 0$, the identity \eqref{Stokes} can be used to write the first term in the integral as
\beq\label{SInt}
\int_\Sigma \nabla_\mu W^\mu k^\alpha d\Sigma_\alpha = 2\int_{\partial\Sigma}W^\mu k^\alpha dS_{\mu\alpha} \ .
\eeq
The surface $\partial\Sigma$ is the 2-surface at infinity and all the terms in $W^\mu$ decay, asymptotically, faster than $r^{-2}$, so that \eqref{SInt} vanishes. Thus~\eqref{mass} becomes
\beq
M = 2 \int_\Sigma U k^\alpha d\Sigma_\alpha = - 2 \int_\Sigma U V d\Sigma \ ,
\eeq
as $d\Sigma_\alpha = k_\alpha d\Sigma$. Consequently, as long as the potential $U(\theta)$ is positive, the only contribution to the Komar mass $M$ will be negative. Then, by the the positive mass theorem\footnote{The energy conditions are unchanged from the Einstein-Maxwell-scalar theory by the axionic term, so we can take the same conclusions as in \cite{HerdeiroOliveira}. The dominant energy condition stays  valid and, as consequence, the positive energy theorem is also valid.},  $M=0$ and the only solution is flat spacetime. Therefore, no axionic solitons are possible in strictly stationary spacetimes, again regardless of the spatial symmetries.

\section{Conclusion}
\label{section5}
In this paper we have assessed the possible existence of self-gravitating solitons in axion electrodynamics and generalisations thereof. We established that the presence of axions and their coupling to the electromagnetic field does not change the results of (in)existence of Einstein-Maxwell solitons in static or strictly stationary spacetime~\cite{HerdeiroOliveira}. This holds even when considering a model with rather generic couplings between the axion field and the electromagnetic invariants, and, in particular allowing an arbitrary coefficient function $g(\theta)$ in the axion  term $F\cdot\tilde{F}$. 

A possible generalisation would be to consider a coupling between the electromagnetic field and a different  scalar field (rather than the axion). However, without any kind of coupling between these two scalar fields, the result will likely remain unchanged.
One interesting future work route would be then to generalize this model to allow for two different scalar fields, coupled to each other, and to the electromagnetic field through the couplings $f$ and $g$. An example of a model that corresponds to this type of framework is the Einstein-Maxwell-dilaton-Axion model\cite{EMDA}, where the coupling $f(\varphi)=e^{-\alpha\varphi}$ depends on the dilaton field $\varphi$ ($\alpha$ is a constant) and $g(\theta)=\kappa\theta$ has the usual dependence on the axion field $\theta$. These two fields also include a coupling between them, possibly allowing for the existence of scalar solitons in the model.

\section*{Acknowledgements}
We would like to thank E. Radu for discussions. J.O. is supported by the FCT grant PD/BD/128184/2016. 
This work has been supported by FCT (Portugal) through: the IF programme, grant PTDC/FIS-OUT/28407/2017, 
the strategic project UID/MAT/04106/2019 (CIDMA) and the CENTRA strategic project UID/FIS/00099/2013. 
We also acknowledge support from  the  European  Union's  Horizon  2020  
research  and  innovation  (RISE) programmes H2020-MSCA-RISE-2015
Grant No.~StronGrHEP-690904 and H2020-MSCA-RISE-2017 Grant No.~FunFiCO-777740. 
The authors would like to acknowledge
networking support by the
COST Action CA16104.

\end{document}